\documentstyle[12pt]{article}
\begin{document}
\def\ve{\varepsilon}
\def\lsim{\:\raisebox{-0.5ex}{$\stackrel{\textstyle<}{\sim}$}\:}
\def\gsim{\:\raisebox{-0.5ex}{$\stackrel{\textstyle>}{\sim}$}\:}
\def\qb {{\bf q}}
\def\pb {{\bf p}}
\def\rb {{\bf r}}
\def\vb {{\bf v}}
\def\Yb{ {\bf Y}}
\def\mba{\mbox{\boldmath$\alpha$}}
\def\mbs{\mbox{\boldmath$\sigma$}}
\def\d{\dagger}
\def\s {{\sigma}}
\def\w {{\omega}}
\def\G {{{\cal G}}}
\def\M {{{\cal M}}}
\def\bin#1#2{\left(\negthinspace\begin{array}{c}#1\\#2\end{array}\right)}
\def\be{\begin{equation}}
\def\ee{\end{equation}}
\def\br{\begin{eqnarray}}
\def\er{\end{eqnarray}}
\def\brn{\begin{eqnarray*}}
\def\ern{\end{eqnarray*}}
\def\x{\times}
\def\fot{\frac{1}{2}}
\def\go{\rightarrow  }
\def\rf#1{{(\ref{#1})}}
\def\bra#1{\langle #1|}
\def\ket#1{|#1 \rangle}
\begin{titlepage}
\pagestyle{empty}
\baselineskip=21pt
\vskip .2in
\begin{flushright}
To be published in Physics Letters B
\end{flushright}
\begin{center}
{\large{\bf Two neutrino double beta decay within the $\xi$-approximation}}
\vskip .1in

C. Barbero$^*$, F. Krmpoti\'{c}$^{\d}$ and A. Mariano$^{\d}$

{\small\it Departamento de F\'\i sica, Facultad de Ciencias Exactas,}\\
{\small\it Universidad Nacional de La Plata, C. C. 67, 1900 La Plata,
Argentina}
\end{center}
\vskip 0.5in
\centerline{ {\bf Abstract} }
\baselineskip=18pt
We examine the contributions of odd-parity nuclear operators to the
two-neutrino double beta decay $0^+\go 0^+$ amplitude, which come from
the $P$-wave Coulomb corrections to the electron wave functions and the recoil
corrections to the nuclear currents. Although they are formally of higher order
in $\alpha Z/2$ or $v/c$ of the nucleon than the usual Fermi and Gamow-Teller
matrix elements, explicit calculations performed within the QRPA show that they
are significant when confronted with the experimental data.

\bigskip

\vspace{0.5in}
\noindent
$^*$Fellow of the Fundaci\'on Antorchas
and of the Universidad Nacional de La Plata.\\
$^{\dagger}$Fellow of the CONICET from Argentina.
\end{titlepage}
\baselineskip=18pt

\newpage
The two neutrino double beta decay ($2\nu\beta\beta$) $0^+\go 0^+$
is the rarest process observed so far in nature. As such, it offers a
unique opportunity for testing the nuclear physics techniques for half-lives
$T_{1/2}^{2\nu} \sim (10^{19}-10^{23})$ years
\cite{Bec93,Ell92,Kaw93,Eji91,Ber92,Art93}.
But, before resorting to a particular nuclear model, the expression for the
$2\nu\beta\beta$ decay rate has to be figured out from the weak Hamiltonian.
\footnote{The $2\nu\beta\beta$-decay is a second order nuclear weak
transition, analogous to the electromagnetic nuclear double gamma decay.
It could be interesting to notice that the half-life for the $\gamma\gamma$
$0^+_2\go 0^+_1$ transition in $^{90}Zr$ is only $T_{1/2}^{2\gamma}
\sim 1.4~\x 10^{-4}$ seconds \cite{Sch84}.}
Except for the work of Williams and  Haxton \cite{Wil88}, the allowed
approximation (AA) has  been universally adopted in all previous theoretical
studies. This approximation  is obtained in
the long-wavelenth limit for the outgoing electrons and
neutrinos by retaining only the Fermi (F) and Gamow-Teller (GT)
operators in the non-relativistic impulse approximation for the hadronic
current
\be
J^{\mu}(\rb)=\sum_n\tau_n^+\left[(g_V-g_A\vb_n\cdot\mbs_n)g^{\mu 0}
+(g_A\s_n^k-g_Vv_n^k)g^{\mu k}\right]\delta(\rb-\rb_n).
\label{1}\ee
Here the operator $\tau^+$ transforms a neutron into a proton and $\mbs_n$,
$\rb_n$ and $\vb_n$ stand, respectively, for the spin, the position and the
velocity of the nucleon $n$. The well known result for the half-live is
\cite{Kon66,Hax84,Doi85,Tom91}
\be
\left[T_{1/2}^{2\nu}(0^+\go 0^+)\right]^{-1}=\G_{2\nu}(W_0,Z)|\M_{2\nu}|^2,
\label{2}\ee
where $\G(W_0,Z)$ is the phase space kinematical factor, and
\br
\M_{2\nu}\equiv\M_{2\nu}^A&=&-g_V^2\sum_N
\frac{\bra{0_F^+}\tau^+\ket{0_N^+} \bra{0_N^+}\tau^+\ket{0_I^+}}
{E_{0_N^+}-E_{0_I^+}+\fot W_0}
\nonumber\\
&+&g_A^2\sum_N\frac{\bra{0_F^+}\tau^+\mbs\ket{1_N^+}\cdot
\bra{1_N^+}\tau^+\mbs\ket{0_I^+}}{E_{1_N^+}-E_{0_I^+}+\fot W_0},
\label{3}\er
is the nuclear matrix element in the AA, with all the notation
having the usual meaning \cite{Hax84,Doi85}.

The retardation of the $0^+\go 0^+$ $2\nu\beta \beta$ transition rates
mostly comes from the phase space factor.
Yet, the nuclear correlations also slow down these processes.
Their effect can be estimated by comparing the measured $|\M_{2\nu}|$ values,
shown in table \ref{tab1}, with the naive estimate, $\M_{2\nu}\sim -0.3$
that is obtained within the closure approximation for an effective axial
charge $g_A^{eff}=1$ \cite{Kon66}. The resulting suppression factor ranges
from $1/2$ for $^{100}Mo$ to $1/15$ for $^{130}Te$.

The quasiparticle random phase approximation (QRPA) is the nuclear structure
method most widely used to deal with the quenching of the moment $\M_{2\nu}$.
We know that in this model the allowed moment $\M_{2\nu}^A$ is very sensitive
to the ground state correlation (GSC) within the particle-particle (PP)
channel. In a recent work \cite{Krm93} it has been shown both that:
(a) the restoration of the isospin symmetry leads to ${\cal M}_{2\nu}^A
(J^{\pi}=0^+) \cong 0$, and (b) the exact QRPA calculations for
${\cal M}_{2\nu}^A(J^{\pi} =1^+)$, when evaluated with a zero-range force,
can be nicely fitted by the formula
\be
{\cal M}_{2\nu}^A(J^{\pi}=1^+)={\cal M}_{2\nu}^A(J^{\pi}=1^+;t=0)
\frac{1-t/t_0}{1-t/t_1},
\label{4}\ee
where $t=v^{pp}_t/v^{pair}_s$ is the ratio between the spin-triplet strength
in the PP and the spin-singlet strength in the pairing channel. The free
parameters $t_0$ and $t_1$ denote the zero and the pole
parameters $t_0$ and $t_1$, with $t_1 \gsim t_0$, denote the zero and the pole
of ${\cal M}_{2\nu}^A$, respectively.
In the same work it has been suggested that this result is of general
validity and that any modification of the nuclear hamiltonian or of the
configuration space cannot lead to a different functional dependence.
Several alterations of the QRPA have been proposed that might change that
behavior. They include higher order RPA corrections \cite{Rad91}, nuclear
deformation \cite{Rad93}, single-particle self-energy BCS terms \cite{Kuo92}
and particle number projection \cite{Krm93a}. Yet, none of these amendments
inhibits the matrix element ${\cal M}_{2\nu}^A$ to pass through zero near
the "natural" value of the PP strength ($t\sim 1.5$).
\footnote{When a physical quantity has a zero (or near zero) a conservative
law should, very likely, be its origin. Thus, the behavior of
${\cal M}_{2\nu}^A$, described by \rf{4}, has been attributed to the partial
restoration of the SU(4) Wigner symmetries \cite{Hir90}. This is not
surprising since the operators $\tau^{\pm}\mbs$ are infinitesimal generators
of the $SU(4)$ group.}

We also know that the GSC do not affect the $0\nu\beta\beta$-decay probability
in the same way. The reason for that is that the exchange of the virtual
neutrino in the $0\nu\beta\beta$ process gives rise to the neutrino potentials,
which depend on the distance $r_{12}=|\rb_1-\rb_2|$ between the two vertices
where the single $\beta$-decays take place. If one expands this dependence on
$r_{12}$ in multipoles depending on $\rb_1$ and $\rb_2$, one sees that,
besides exciting F and GT states, one is also going through virtual states
with spin and parity $J^{\pi}\neq 0^+,~1^+$. The GSC are not so important for
the forbidden transitions $(L>0)$ and thus they are not so drastically reducing
the decay rate for the $0\nu\beta\beta$-decay as for the
$2\nu\beta\beta$-decay.
In fact, it turns out that for $t\sim 1.5$ the dominant contribution to
${\cal M}_{0\nu}$ comes from multipoles with $L>0$ \cite{Krm93}.

Motivated by the above arguments and by the hope that some
favorable nuclear structure physics could enhance the higher order
contributions to ${\cal M}_{2\nu}$, we study here the effects of the
$2\nu\beta\beta$ matrix elements with $L=1$.
It will be assumed that the Coulomb energy of the electron at the nuclear
radius is larger that its total energy. This leads to the $\xi$-approximation,
which has been extensively utilized in the study of the first forbidden (FF)
single $\beta$ transitions \cite{Beh82}. Besides, it is well known that, when
this approximation is applied, the FF observables can be expressed exactly as
the allowed ones except for the replacements \cite{Kon66}

\be
\tau^{+}g_V\go-\tau^{+}X;\hspace{2cm}
\tau^{+}g_A\mbs\go -\tau^{+}\Yb,
\label{5}\ee
with
\br
X&=&g_A(\mbs\cdot\vb+\xi i\mbs\cdot\rb),
\nonumber\\
\Yb&=&g_V\vb+\xi(g_Vi\rb -g_A\mbs\x\rb),
\label{6}\er
where (in natural units)
\be
\xi=\frac{\alpha Z}{2R}\cong 1.18~Z A^{-1/3},
\label{7}\ee
$Z$ and $R$ being the charge and the radius of the daughter nucleus.
\footnote{We discuss here only the $\beta^-$-decay.
For positron emission the following changes are made: $Z\go -Z$
and $g_A\go -g_A$.}
This result is obtained by retaining the velocity dependent terms in the
hadronic
current \rf{1} and by considering only: i) the $S-$wave for the neutrino wave
function $\phi^{\nu}(s)$, and ii) the $S-$ wave and the $\xi$ correction
term in the $P-$wave for the electron wave function $\psi^e(\rb,s)$, i.e.,
\be
\phi^{\nu}(s)=
\frac{1}{\sqrt{2}}\bin{I}{(\mbs\cdot\hat{\qb})}\chi^{\nu}(s)
\label{8}\ee
and
\be
\psi^e(\rb,s)=\sqrt{\frac{F_0(Z,\ve)(\ve+m)}{2\ve}}
\left[1+i\xi(\mba\cdot\rb)\right]\bin{I}
{\frac{(\mbs\cdot\pb)}{\ve+m}}\chi^e(s),
\label{9}\ee
where $F_0(Z,\ve)$ is the Fermi factor and all the remaining notation
is that of ref. \cite{Doi85}.

Before proceeding let us recall that the contributions of all five FF moments
are of the same order of magnitude. First, in the extreme single particle
model
\be
\frac{\bra{f}\mbs\x\rb\ket{i}}
{\bra{f}i\rb\ket{i}} =j_f(j_f+1)-l_f(l_f+1) -j_i(j_i+1)+l_i(l_i+1).
\label{10}\ee
Next, the velocity dependent $\beta$-moments are related in a simple way to
the corresponding position operators, when the harmonic oscillator radial wave
functions are employed. One gets \cite{Beh82,War88,War94}:
\be
\frac{\bra{f}\mbs\cdot\vb\ket{i}}{\bra{f}i\mbs\cdot\rb\ket{i}}
=\frac{\bra{f}\vb\ket{i}}{\bra{f}i\rb\ket{i}}
=(N_f-N_i)\w_0,
\label{11}\ee
where $N$ and $\w_0\cong 80~A^{-1/3}$ stand for the principal quantum number
and the oscillator frequency, respectively.
The most relevant single-particle transitions, in any nuclear model
calculation, are those with $N_i=N_f+1$ (see e.g., table 2 in ref.
\cite{War88}).
\footnote{Note that the operators $i\mbs\cdot\rb$ ($i\rb$) and $\mbs\cdot\vb$
($\vb$, $\mbs\x\rb$) have different Hermitian conjugation properties
[expressed in equations \rf{10} and \rf{11}].
This makes that in a RPA calculation the signs of the ratios \rf{10} and
\rf{11} are different for the forward and backward going contributions.}
Consequently the contributions of the operators $\mbs\cdot\vb$ and
$\vb$ always add destructively to those coming from the operators of
$i\xi\mbs\cdot\rb$ and $i\xi\rb$ (see eq. \rf{6}). In addition, the ratio of
the corresponding matrix elements is $\approx -68/Z$, i.e., of order of unity
for medium heavy nuclei.

The conserved-vector-current (CVC) constraint leads to an alternative
expression for the total matrix element $<\vb>$ \cite{Bli66}
\be
<\vb>=-(2.4\xi+E_0)<i\rb>,
\label{12}\ee
where $E_0$ is the energy difference between the initial and final states.
(In principle, this relation is exact as it includes exchange effects, induced
interactions, etc. \cite{Beh82}.)
Besides, it is now well established \cite{Ade86} that the operator
$\mbs\cdot\vb$ is enhanced by the one-pion exchange current contributions in
nuclear medium \cite{Kub78}. Inspired by this effect, which can be accounted
for by renormalizing the corresponding matrix element by the factor
$1.7~g_A^{free}\cong 2.0$ (see also ref. \cite{War94}), Williams and Haxton
\cite{Wil88} have pointed out that the moment $<\mbs\cdot\vb>$ could have
a significant effect on the rates for $2\nu\beta\beta$-decay. They also
performed calculations in the Nilsson pairing model with the
quadrupole-quadrupole core-polarization effect included.
The result was to increase the AA predictions for
$T_{1/2}^{2\nu}$ of $^{76}Ge$, $^{82}Se$, $^{128}Te$ and $^{130}Te$ nuclei by
$15\%$, $25\%$, $15\%$ and $20\%$, respectively (see also ref. \cite{Hax92}).
Yet from the above discussion it is clear that the consequences of the
remaining four FF moments have also be examined simultaneously.
\footnote{Williams and  Haxton \cite{Wil88} have also hinted that, besides
$<\mbs\cdot\vb>$, other parity-forbidden matrix elements should be considered.
Particularly, because of the CVC relation, they have pointed towards the
importance of $<\vb>$.}
Such an examination is the main topic of this paper.

The expression for the half-life for the $0^+\go 0^+$ $2\nu\beta\beta$-decay,
within the $\xi$-approximation, has been derived following the procedure
of Doi et al. \cite{Doi85} and using the relations \rf{8} and \rf{9}
for the lepton wave functions. A tedious  but straightforward calculation
\cite{Bar94} leads to the equation for the half-life that can again be cast
in the form \rf{1}, with the {\em same kinematical factor} and the transition
amplitude
\be
\M_{2\nu}\equiv\M_{2\nu}^{\xi}= \M_{2\nu}^A+ \M_{2\nu}^{FF},
\label{13}\ee
where
\br
\M_{2\nu}^{FF}=&
-&\sum_N\frac{\bra{0_F^+}\tau^+X\ket{0_N^-}
\bra{0_N^-}\tau^+X\ket{0_I^+}}{E_{0_N^-}-E_{0_I^+}+\fot W_0}
\nonumber\\
&+&\sum_N\frac{\bra{0_F^+}\tau^+\Yb\ket{1_N^-}\cdot
\bra{1_N^-}\tau^+\Yb\ket{0_I^+}}{E_{1_N^-}-E_{0_I^+}+\fot W_0},
\label{14}\er
is the contribution coming from the FF moments. It is also important to note
that the shape of the electron spectrum is the same as in the AA.

As the spin singlet components predominate in the
ground state of the even-even nuclei, the two terms in \rf{3} and \rf{14}
have a tendency to sum up  coherently.
On the other hand, it is self-evident from relations \rf{6} that,
independently of the nuclear model employed in the evaluation of
$\M_{2\nu}^{FF}$, the role played by the virtual states with $J^{\pi}=0^-$
and $J^{\pi}=1^-$ critically depend on the way in which the individual matrix
elements combine to build up the total matrix elements of $X$ and $\Yb$.
Coherent combinations would strongly enhance the effect of
$\M_{2\nu}^{FF}$. Yet, from the above discussion we have learned that
the moments $<\mbs\cdot\vb>$ and $<i\mbs\cdot\rb>$, as well as $<\vb>$
and $<i\rb>$, always tend to add destructively.
Hence the magnitude of $\M_{2\nu}^{FF}$ hangs crucially on the
matrix element $<\mbs\x\rb>$ (both on its magnitude and its
sign relative to the sign of $<\vb>$).
\footnote{It is worth noting that $\M_{2\nu}^{FF}$ can be derived from
$\M_{2\nu}^{A}$ by performing the substitution \rf{5} in \rf{3},
i.e., in the analogous  way that are obtained the FF observables from the
allowed ones in the simple $\beta$ decay. This common future of the single
and double beta decays came somewhat as a surprise.
One should remember here that, because of the antisymmetrization in the
momentum and spin-orientation, the $2\nu\beta\beta$ amplitude is not just
a simple product of two single $\beta$ decays. Yet, a careful examination
of the analytic calculations shows that it is precisely the
anti-symmetrization procedure that establishes such a correspondence.}

The numerical results for the matrix elements $\M_{2\nu}^{FF}$
are presented in table \ref{tab2}.
They were obtained through the procedure adopted in our previous works
\cite{Hir90}, i.e., by using the same residual interaction, the same
configuration space, etc.
The moments labeled ${\cal M}_{2\nu}^{FF}(pair)$ were evaluated in the
independent quasiparticle approximation, and those indicated by
${\cal M}_{2\nu}^{FF}(QRPA)$ within the QRPA. The matrix element
$<\mbs\cdot\vb>$ has been estimated by using the relation \rf{11},
properly renormalized by the one-pion exchange currents. The CVC relation
\rf{12} has been used in the evaluation of the moment $<\vb>$.
The particle-hole interaction reduces the forbidden moments by roughly a
factor of $2$. But, at variance with the moments ${\cal M}_{2\nu}^A$, which
are widely uncertain in the QRPA (due to the proximity of $t_1$ to $t_0$
in eq. \rf{4}), the effect of the particle-particle interaction on
${\cal M}_{2\nu}^A$ is always very small.
For all the nuclei, the contributions of the virtual states with $J^{\pi}=0^-$,
shown parenthetically in table \ref{tab2}, are smaller than those coming from
the states with $J^{\pi}=1^-$. There are two reasons for that. First, the
individual single-particle matrix elements of the operator $\Yb$ are
usually larger than those of the operator $X$. This is caused mainly
by the destructive interference between the individual FF matrix elements
mentioned above.
Among the former the most important are those with $\bra{f}\mbs\x\rb\ket{i}
=\bra{f}i\rb\ket{i}$, in which case $\bra{f}\Yb\ket{i}=\bra{f}\vb\ket{i}$
(for $g_A^{eff}=1$). Second, for both type of transitions the partial
contributions show a pronounced coherence, but the configuration space for
the $1^-$ is bigger. For the sake of comparison, the matrix elements
$\M_{2\nu}^{A}$ are displayed in table 2 also.

We note that the calculation done by Williams and Haxton \cite{Wil88}
yields considerably larger contributions of the $J^{\pi}=0^-$ states
that our QRPA evaluation. For instance, for the $^{82}Se\go~^{82}Kr$ and
$^{130}Te\go~^{130}Xe$ transitions they get $\M_{2\nu}^{FF}(0^-) \cong 0.010$.
The discrepancy with our results can be attributed only partially to the effect
of the matrix element $~<i\mbs\cdot\rb>$, which has not been considered in ref.
\cite{Wil88}. By neglecting the last one we obtain that $\M_{2\nu}^{FF}
(0^-)\cong0.006$ for both nuclei. The remaining discrepancy has to be attached
to the difference in the nuclear models employed in ref. \cite{Wil88} and here.

{}From the results shown in table \ref{tab2} it is clear that in the QRPA the
matrix elements $\M_{2\nu}^{FF}$ are significant when compared with
the $\M_{2\nu}^{A}$ moments. But, the real relevance of $\M_{2\nu}^{FF}$ comes
from the confrontation with the experimental data. In fact, we see that they
are roughly comparable to $|{\cal M}_{2\nu}|(exp)$ needed to explain the half
lives of $^{128}Te$, $^{130}Te$, and $^{150}Nd$ nuclei.
For the remaining four nuclei displayed in tables \ref{tab1} and \ref{tab2}
the moments $\M_{2\nu}^{FF}$ are relatively small in comparison with the
measured amplitudes. Yet, even in this case the parity forbidden moments
might become significant in the theoretical evaluation of the
$2\nu\beta\beta$-decay half-lives, as they have a tendency to cancel against
the allowed moments. Besides, the sensitivity of $2\nu\beta\beta$ experiments
has been greatly improved in recent years, and there is no reason to believe
that further improvements are not still possible.

In closing, we point out that there are other forbidden terms that should be
studied. For instance, we are in process of examining the contributions
arising from virtual states $J^{\pi}=2^-$, which might modify both the
electron spectrum shape and the $\G(W_0,Z)$ factor predicted by the AA.
This could be of relevance for the excess of high-energy electrons seen
in the $\beta\beta$ spectra and presently ascribed to the majoron emission.
Finally, we notice that in the new class of majoron models the
$\beta\beta$-decay proceeds via the odd-parity nuclear operators \cite{Bur93}.

We warmly thank Dubravko Tadi\'c for fruitful discussions.
%

\newpage
\begin{table}[t]
\begin{center} {\large Tables }
\caption {The measured half-lives $T_{1/2}^{2\nu}$ and the phase space
function $G_{2\nu}(W_0,Z)$, which were used to extract the values of
$|{\cal M}_{2\nu}|(exp)$, given in natural units.}
\label{tab1}
\bigskip
\begin{tabular}{l|ccc}
\hline
\hline
Nucleus&${T_{1/2}^{2\nu} [yr \; 10^{20}]}$
&$\!G^{2\nu}[yr]^{-1}\!$
&$|{\cal M}_{2\nu}|(exp)$\\
\hline
{$^{76}Ge$}&$\!14.3 \pm 0.17\,^{a)}\!$&$\!5.39\;10^{-20}\!$
&$0.114\pm 0.007$\\
{$^{82}Se$}&$\!1.08_{-0.06}^{+0.26}\,^{b)}\!$&$\!1.80\;10^{-18}\!$
&$0.071_{-0.007}^{+0.002}$\\
{$^{96}Zr$}& $\!0.39\pm0.09\,^{c)}\!$ &$\!7.66\;10^{-18}\!$
&$0.058_{-0.006}^{+0.008}$\\
{$^{100}Mo$}&$\!0.115_{-0.020}^{+0.030}\,^{d)}\!$&$\!3.91\;10^{-18}\!$
&$0.149_{-0.017}^{+0.015}$\\
{$^{128}Te$}&$\!(7.7\pm0.4)\;10^{4}\,^{e)}\!$&$\!3.53\;10^{-22}\!$
&$0.019\pm 0.001$\\
{$^{130}Te$}&$\!27\pm1\,^{e)}\!$ &$\!1.98\; 10^{-18}\!$&$0.014\pm 0.001$\\
{$^{150}Nd$}&$\!0.17_{-0.09}^{+0.14}\,^{f)}\!$&$\!4.91\;10^{-17}\!$
&$0.035_{-0.009}^{+0.014} $\\
\hline \hline \end{tabular} \end{center}
$^{a})$ (laboratory data) ref.\ \cite{Bec93}\\
$^{b})$ (laboratory data) ref.\ \cite{Ell92}\\
$^{c})$ (geochemical data) ref.\ \cite{Kaw93}\\
$^{d})$ (laboratory data) ref.\ \cite{Eji91}\\
$^{e})$ (geochemical data) ref.\ \cite{Ber92}\\
$^{f})$ (laboratory data) ref.\ \cite{Art93}\\
\end{table}
\bigskip
\begin{table}[h]
\begin{center}
\caption {Calculated allowed and first forbidden matrix elements (in natural
units) for an effective axial charge $g_A^{eff}=1$.
The column labeled ${\cal M}_{2\nu}^{FF}(pair)$ is evaluated using the
independent quasiparticle approximation, while the column
${\cal M}_{2\nu}^{FF}(QRPA)$ includes the effects of both the particle-hole
and the particle-particle interactions.
The values in parenthesis stand for the contributions of virtual states with
$J^{\pi}=0^-$ to the total moments.}
\label{tab2}
\bigskip
\begin{tabular}{l|cccc|c} \hline \hline
\multicolumn{1}{c|}{Nucleus}&\multicolumn{2}{c}{${\cal M}_{2\nu}^{FF}(pair)$}
&\multicolumn{2}{c|}{${\cal M}_{2\nu}^{FF}(QRPA)$}
&\multicolumn{1}{c}{${\cal M}_{2\nu}^A(QRPA)$}\\
\hline
{$^{76}Ge$} &$-0.015$&$(-0.005)$&$-0.008$&$(-0.004)$&$0.050$\\
{$^{82}Se$} &$-0.019$&$(-0.006)$&$-0.009$&$(-0.004)$&$0.060$\\
{$^{96}Zr$} &$-0.033$&$(-0.010)$&$-0.014$&$(-0.006)$&$0.010$\\
{$^{100}Mo$}&$-0.035$&$(-0.010)$&$-0.014$&$(-0.006)$&$0.051$\\
{$^{128}Te$}&$-0.020$&$(-0.004)$&$-0.012$&$(-0.003)$&$0.059$\\
{$^{130}Te$}&$-0.019$&$(-0.004)$&$-0.012$&$(-0.002)$&$0.048$\\
{$^{150}Nd$}&$-0.060$&$(-0.009)$&$-0.031$&$(-0.008)$&$0.033$\\
\hline \hline \end{tabular} \end{center}\end{table} 
\newpage
\begin{thebibliography}{99}
\bibitem{Bec93}  M. Beck et al., Phys. Rev. Lett. {\bf 70}  (1993) 2853
\bibitem{Ell92}  S.R. Elliot, A.A. Hahn, M.K. Moe, M.A. Nelson and M.A. Vient,
Phys. Rev. {\bf C46}  (1992) 1535
\bibitem{Kaw93}  A. Kawashima, K. Takahashi and A. Masuda,
Phys. Rev. {\bf C47}  (1993) 2452
\bibitem{Eji91} H. Ejiri, K. Fushimi, T. Kamada, H. Kinoshita, H. Kobiki, H.
Ohsumi,
K. Okada, H. Sano, T. Shibata, T. Shima, N. Tanabe, J. Tanaka, T. Taniguchi,
T. Watanabe and N. Yamamoto, Phys. Lett. {\bf B258} (1991) 17
\bibitem{Ber92} T. Bernatowitz, J. Brannon, R. Brazlle, R. Cowsik, C.
Hohenberg,
and F. Podosek, Phys. Rev. Lett. {\bf 69} (1992) 2341;
Phys. Rev. {\bf C47} (1993) 806
\bibitem{Art93} V.A. Artem'ev et al., JETP Lett. {\bf 58} (1993) 262.
\bibitem{Sch84} J. Schimer, D. Habs, R. Kroth, N. Kwong, D. Schwalm and
M. Zirnbauer, Phys. Rev. Lett. {\bf 53} (1984) 1897;
L.P. Ekstr\"{o}m and J. Lyttkens-Linden, Nucl. Data Sheets {\bf 67} (1992) 579.
\bibitem{Wil88} A. Williams and W.C. Haxton, in {\it Intersections between
Particle and Nuclear Physics}, ed. G.M. Bunce (AIP Conf. Proc. No. 176, 1988)
p. 924.
\bibitem{Kon66} E.J. Konopinski, {\it Theory of Beta Radioactivity}
(London: Oxford University Press, 1966).
\bibitem{Hax84} W.C. Haxton and G. J. Jr. Stephenson, Prog. Part. Nucl. Phys.
{\bf 12} (1984) 409.
\bibitem{Doi85} M. Doi, T. Kotani and E. Takasugi, Prog. Theor. Phys. Suppl.
{\bf 83} (1985) 1; Phys. Rev. {\bf D37} (1988) 2104.
\bibitem{Tom91} T. Tomoda, Rep. Prog. Phys. {\bf 54} (1991) 53.
\bibitem{Krm93} F. Krmpoti\'{c}, Phys. Rev. {\bf C48} (1993) 1452.
\bibitem{Rad91} A.A. Raduta, A. Faessler, S. Stoica and W.A. Kaminski,
Phys. Lett. {\bf B254} (1991) 7.
\bibitem{Rad93} A.A. Raduta, A. Faessler and D.S. Delion,
Nucl. Phys. {\bf A564} (1993) 185.
\bibitem{Kuo92} D.B. Stout and T.T.S. Kuo, Phys. Rev. Lett. {\bf 69} (1992)
1900; S.S. Hsiao, Yiharn Tzeng and T.T.S. Kuo, Phys. Rev. {\bf C49} (1994)
2233.
\bibitem{Krm93a} F. Krmpoti\'{c}, A. Mariano, T.T.S. Kuo and K. Nakayama,
Phys. Lett. {\bf B319} (1993) 393.
\bibitem{Hir90} J. Hirsch and F. Krmpoti\'{c}, Phys. Rev. {\bf C41} (1990) 792;
J. Hirsch, E. Bauer and F. Krmpoti\'{c}, Nucl. Phys. {\bf A516} (1990) 304;
F. Krmpoti\'{c} and S. Shelly Sharma, Nucl. Phys. {\bf A572} (1994) 329.
\bibitem{Beh82} H. Behrens and W. B\"uring, {\it Electron Radial Wave
Functions and Nuclear Beta Decay} (Clarendon, Oxford, England, 1982).
\bibitem{War88} E.K. Warburton, J.A.Becker, B.A. Brown and D.J. Millener,
Ann. Phys. (N.Y.) {\bf 187} (1988) 471.
\bibitem{War94} E.K. Warburton, I.S. Towner and B.A. Brown, Phys. Rev.
{\bf C49} (1994) 824.
\bibitem{Bli66} R.J. Blin-Stoyle and S.C.K. Nair, Adv. Phys. {\bf 15} (1966)
493; R.J. Blin-Stoyle, {\it Fundamental Interactions and the Nucleus}
(North Holland, Amsterdam, 1973).
\bibitem{Ade86} E.G. Adelberger and W.C. Haxton, Ann. Rev. Nucl. Part. Sci.
{\bf 35} (1986) 501; I.S. Towner, {\it ibid.} {\bf 36} (1986) 115.
\bibitem{Kub78} K. Kubedera, J. Delorme and M. Rho, Phys. Rev. Lett. {\bf 40}
(1978) 755.
\bibitem{Hax92} W.C. Haxton, Nucl. Phys. (Proc. Suppl.) {\bf 31} (1993) 88.
\bibitem{Bar94} C. Barbero, M. Sc. Thesis, Universidad Nacional de La Plata,
1994 (unpublished).
\bibitem{Bur93} C.P. Burgess and J.M. Cline, Phys. Lett. {\bf B298} (1993) 141;
Phys. Rev. {\bf D49} (1994) 5925, and references therein.
\end{thebibliography}
\end{document}